\documentclass{elsart}

\usepackage{graphicx}

\usepackage{amssymb, amsmath}

\begin{document}
\newcommand{\Tr}{{\rm Tr}}
\begin{frontmatter}



\title{Full replica symmetry breaking in generalized
mean--field spin glasses with reflection symmetry}


\author{ T.I.Schelkacheva,
E.E.Tareyeva, \and N.M.Chtchelkachev}

\address{Institute for High Pressure Physics,
Russian Academy of Sciences, Troitsk 142190, Moscow region,
Russia}

\begin{abstract}
 The analysis of the solution with full replica symmetry breaking in the
 vicinity of $T_c$ of a general spin glass model with reflection symmetry
 is performed. The leading term in the order parameter
 function expansion is obtained. Parisi equation for the model is written.
\end{abstract}

\end{frontmatter}



The crucial role of the reflection symmetry for the character of
phase transition in nonrandom mean--field (MF) models is well
known (see, e.g., the textbook ~\cite{LLSt}). Generally speaking
the presence of the terms without reflection symmetry causes the
first order phase transition, while in the absence of such terms
the transition is of the second order. Usually this result is
obtained in the frame of the phenomenological approach based on
the Ginzburg--Landau (GL) effective Hamiltonian that can be
easily obtained for any Hamiltonian through the
Hubbard--Stratanovich identity for the partition function.

In the case of random MF models the behavior of systems with
reflection symmetry is quite different from those without it.
In Ref.~\cite{4aut} this problem was investigated in the frame of
replica symmetric (RS) approach. The role of the reflection
symmetry for the behavior of RS solution for spin--glass--like MF systems
was considered and a kind of symmetry rule for the type of the
growing of glass order was formulated: if in the nonrandom (pure) system
the transition to the ordered phase is of the second order, then in the
corresponding random system the glass regime appears as a result of a
phase transition; if the transition in the pure system is of the first
order, then in the random system the glass order parameter grows
continuously on cooling.  In fact, one can imagine that both the first
order phase transition in nonrandom systems as well as the continuous
growing of the glass order in random systems are caused by some kind of
internal fields appearing due to the algebra of operators $\hat U$. In
this case we have a close analogy with Sherrington--Kirkpatrick (SK) model
in an external field.

The scenarios of replica symmetry breaking (RSB) for the mentioned two
cases differ essentially one from another because
the absence of reflection symmetry
results in a special form of RSB free energy functional for
random case, too. So, the absence of reflection symmetry can lead to
continuous growing of the Edwards--Anderson order parameter as well as
to a jump at the transition to one stage replica symmetry breaking (1RSB)
phase and other specific features
(see, for example,
Refs.~\cite{goeld,rev3,crile,book}). In the reflection
symmetrical random systems the glass regime appears as a result of a
bifurcation from the trivial solution for the order--parameter function.
The prototype model is the well known Sherrington--Kirkpatrick model
~\cite{sk}, the fully connected Ising spin model with quenched random
exchange magnetic interactions. This model is described by a full replica
symmetry breaking (FRSB) Parisi solution
 ~\cite{par1,par2,book,Moskalenko}.

Although the features of SK model have been thoroughly investigated ,
comparatively little attention has been paid to the analysis of natural
generalizations of SK model (for example the systems of higher spins).
So, we believe that it would be quite useful to consider a generalized spin
glass model where Ising spins are replaced by arbitrary diagonal operators
$\hat U$ with $\Tr \hat U = 0$. In RS approximation this model was
considered in ~\cite{4aut}. In the present paper we continue the
investigation of this model and consider in detail replica symmetry
breaking in the case when the system possesses reflection symmetry.

Let us consider a system of particles on lattice sites $i, j$ with
Hamiltonian
\begin{equation}
H=-\frac{1}{2}\sum_{i\neq j}J_{ij}
\hat{U_i}\hat{U_j}, \label{one}
\end{equation}
where $\hat{U}$ is an arbitrary diagonal operator with
$\Tr\hat{ U}=0$, $J_{ij}$ are quenched Gaussian interactions
 with zero mean
\begin{equation} P(J_{ij})=\frac{1}{\sqrt{2\pi
J}}\exp\left[-\frac{(J_{ij})^{2}}{
2J^{2}}\right] \label{two}
\end{equation}
with $ J=\tilde{J}/\sqrt{N}$.

Using replica approach (see, e.g. ~\cite{sk}) we can write
the free energy averaged over disorder in the form ~\cite{4aut}
\begin{eqnarray}
&&\langle F\rangle_J/NkT=\lim_{n \rightarrow 0}\frac{1}{n}\max\biggl
\{
\frac{t^2}{4}\sum_{\alpha} (p^{\alpha})^{2} +
\frac{t^2}{2}\sum_{\alpha>\beta} (q^{\alpha\beta})^{2}- \nonumber
\\
&&-\ln\Tr_{\{U^{\alpha}\}}\exp\left[
\frac{t^2}{2}
 \sum_{\alpha}p^{\alpha}(\hat{U}^{\alpha})^2+t^2
\sum_{\alpha>\beta}q^{\alpha\beta}\hat{U}^{\alpha}\hat{U}^{\beta}\right]\biggr\}.
\label{free}
\end{eqnarray}
Here $t=\tilde{J}/kT$. The saddle point conditions for the free energy
give the glass order parameter
\begin{equation}
q^{\alpha\beta}=
\frac{\Tr\left[\hat{U}^{\alpha}\hat{U}^{\beta}
\exp\left(\hat{\theta}\right)\right]}
{\Tr\left[\exp\left(\hat{\theta}\right)\right]}
\label{four}
\end{equation}
and auxiliary order parameter
\begin{equation}
 p^{\alpha}=
\frac{\Tr\left[(\hat{U}^{\alpha})^2
\exp\left(\hat{\theta}\right)\right]}
{\Tr\left[\exp\left(\hat{\theta}\right)\right]}
\label{five}
\end{equation}
Here
\begin{equation}
\hat{\theta}=\frac{t^2}{2}
 \sum_{\alpha}p^{\alpha}(\hat{U}^{\alpha})^2+t^2
\sum_{\alpha>\beta}q^{\alpha\beta}\hat{U}^{\alpha}\hat{U}^{\beta}.
\label{six}
 \end{equation}
As one index quantity are usually replica invariant (see e.g.
~\cite{crile} and ~\cite{book})
 we put $p^{\alpha}=p$ from the very
 beginning.

In this paper we study the model (\ref{one}) for operators $\hat{U}$
with zero trace for all odd powers:
\begin{equation}
\Tr\left[\hat{U}^{(2k+1)}\right]=0
\label{seven}
\end{equation}
for all integer $k$.
      In the replica symmetric (RS) approximation the free energy
(\ref{free}) has the
form:
\begin{equation}
F=-NkT\left\{
 t^2\frac{q^2}{4}-t^2\frac{p^2}{4}+
\int_{-\infty}^{\infty}\frac{dz}{\sqrt{2\pi}}\exp\left(-\frac{z^2}{2}\right)\ln
\Tr\left[exp\left(\hat{\theta}\right)\right]\right\}. \label{frs}
\end{equation}

 Here
$$\hat{\theta}_{RS}=zt\sqrt{q_{RS}}\hat{U}+t^2\frac{p-q_{RS}}{2}\hat{U}^2.$$

The extremum conditions for the
free energy ~(\ref{frs}) give the following equations for the
glass order parameter:
\begin{equation}
q_{RS}=\int dz^G\left\{
\frac{\Tr\left[\hat{U}
\exp\left(\hat{\theta}_{RS}\right)\right]}
{\Tr\left[\exp\left(\hat{\theta}_{RS}\right)\right]}\right\}^{2}
\label{qrs}
\end{equation}
and an auxiliary equation
\begin{equation}
p=\int dz^G
\frac{\Tr\left[\hat{U}^2 \exp\left(\hat{\theta}_{RS}\right)\right]}
{\Tr\left[\exp\left(\hat{\theta}_{RS}\right)\right]} \label{prs}
\end{equation}
Here
$$\int dz^G =
\int_{-\infty}^{\infty}
\frac{dz}{\sqrt{2\pi}}\exp\left(-\frac{z^2}{2}\right).$$

If the condition ~(\ref{seven}) is fulfilled then Eq.~(\ref{qrs}) has
a trivial solution for the glass order parameter $q_{RS}=0$ at any
temperature. The nontrivial solution appears at the point $T_c$ defined by
the bifurcation condition  for the Eq.~(\ref{qrs}):
\begin{equation}
1-t_c^2 p(t_c) = 0.
\label{ten}
\end{equation}
The replica symmetric solution is stable unless the replicon
mode energy $\lambda$ is nonzero \cite{almth}. For our model
 ~(\ref{one})-~(\ref{free}) we have:
\begin{equation}
\lambda_{repl(RS)} = 1 -  t^2 \int dz^G
\left\{\frac{\Tr\left[\hat{U}^2
\exp\left(\hat{\theta}_{RS}\right)\right]}
{\Tr\left[\exp\left(\hat{\theta}_{RS}\right)\right]}-
\left[\frac{\Tr\left[\hat{U}
\exp\left(\hat{\theta}_{RS}\right)\right]}
{\Tr\left[\exp\left(\hat{\theta}_{RS}\right)\right]}\right]^2\right\}^2.
\label{lambda}
\end{equation}

In our case ~(\ref{seven}) the condition $\lambda_{repl(RS)} = 0$ is equal to the equation
~(\ref{ten}) due to the fact that it is the trivial solution that bifurcates. Performing one
stage replica symmety breaking (1RSB), then 2RSB, 3RSB etc. we can see that the r.h.s. of the
equations for the glass order parameters always contains $\frac{\Tr[\hat{U} \exp (\hat
{\theta}_{nRSB})]}{\Tr[\exp (\hat{\theta}_{nRSB})]}$. Therefore at each stage of RSB one
possible solution is a trivial one and we can consider the appearance of nRSB solution as the
bifurcation from the trivial (nRSB) solution. In such a case the equation $\lambda _{(nRSB)} =
0$ coincides with the corresponding bifurcation condition ~(\ref{ten}). For example, in the
case of 2RSB we have
$$
\lambda_{repl(1RSB)} = 1 - $$ $$ t^2 \int dz^G \frac{1}
{\int dx^G \left(\Tr[\exp
 \hat{\theta}_{1RSB}(z,x)]\right)^{m}}
\int dy^G \left(\Tr[\exp
 \hat{\theta}_{1RSB}(z,y)]\right)^{m}$$
\begin{equation}
  \left\{\frac{\Tr\left[\hat{U}^2
\exp\hat{\theta}_{1RSB}(z,y)\right]}
{\Tr\left[\exp\hat{\theta}_{1RSB}(z,y)\right]}-
\left[\frac{\Tr\left[\hat{U}
\exp\hat{\theta}_{1RSB}(z,y)\right]}
{\Tr\left[\exp\hat{\theta}_{1RSB}(z,y)\right]}\right]^2\right\}^2.
\label{lambda1}
\end{equation}
Here $$\hat{\theta}_{1RSB}(z,x)=(\sqrt{r}tz + \sqrt
{v}tx)\hat{U} + \frac{t^2}{2}(p-r-v)\hat {U}^2. $$
 The 1RSB solution is
characterized by $m$ and two values $r$ and $r+v$ of $q^{\alpha \beta }$.
 For $r=0$ and $v=0$ the equation
~(\ref{lambda1})
reduces to
 Eq.~(\ref{ten}).

 This means that the RSB solutions of different stages at least may
 exist at temperature $T<T_c$ defined by the condition ~(\ref{ten}) and
 encourage us to look for the full replica symmetry breaking (FRSB)
 ~\cite{par1,par2}.

As an example let us consider the Hamiltonian ~(\ref{one}) with $\hat{U}=\hat{Q}+\eta\hat{V}$,
$\hat{Q}=3\hat{J_z}^2-J(J+1)$, $\hat{V}=\sqrt{3}({\hat{J_x}}^2-{\hat{J_y}}^2)$. The model
describes random quadrupole interaction. This model has been considered for $J=1, J_z = 1, 0,
-1$ in Ref.~\cite{4aut} and for $J=2$ in ~\cite{qg2} in RS approximation.  Here we restrict
ourselves to the case $J=1$. It easy to show that ${\hat{Q}}^2=2-\hat{Q}$,
${\hat{V}}^2=2+\hat{Q}$, $\hat{Q}\hat{V}=\hat{V}\hat{Q}=\hat{V}$ and
\begin{equation}
(\hat{Q}+\eta\hat{V})^2=2(1+\eta^2)+2\eta\hat{V}-(1-\eta^2)\hat{Q}.
\label{alg}
\end{equation}
$$(\hat{Q}+\eta\hat{V})^3=2(3\eta^2-1)+3(1+\eta^2)(\hat{Q}+\eta\hat{V})];$$
$$\Tr(\hat{Q}+\eta\hat{V})^3 = 6(3\eta^2-1).$$
Here $\eta>0$ is a tuning
parameter. The behavior of the system is quite different in the cases
$\eta =1/\sqrt 3$ (analog of spin--one glass model ~\cite{spin1}) and $\eta
\neq 1/\sqrt 3$.

In figures we present the behavior of the 1RSB order parameters and $\lambda_{repl(1RSB)}$ for
this case ($\eta =1/\sqrt 3$) and for  the spin--two glass model.
\begin{figure}[t]
\begin{center}
\includegraphics[height=100mm]{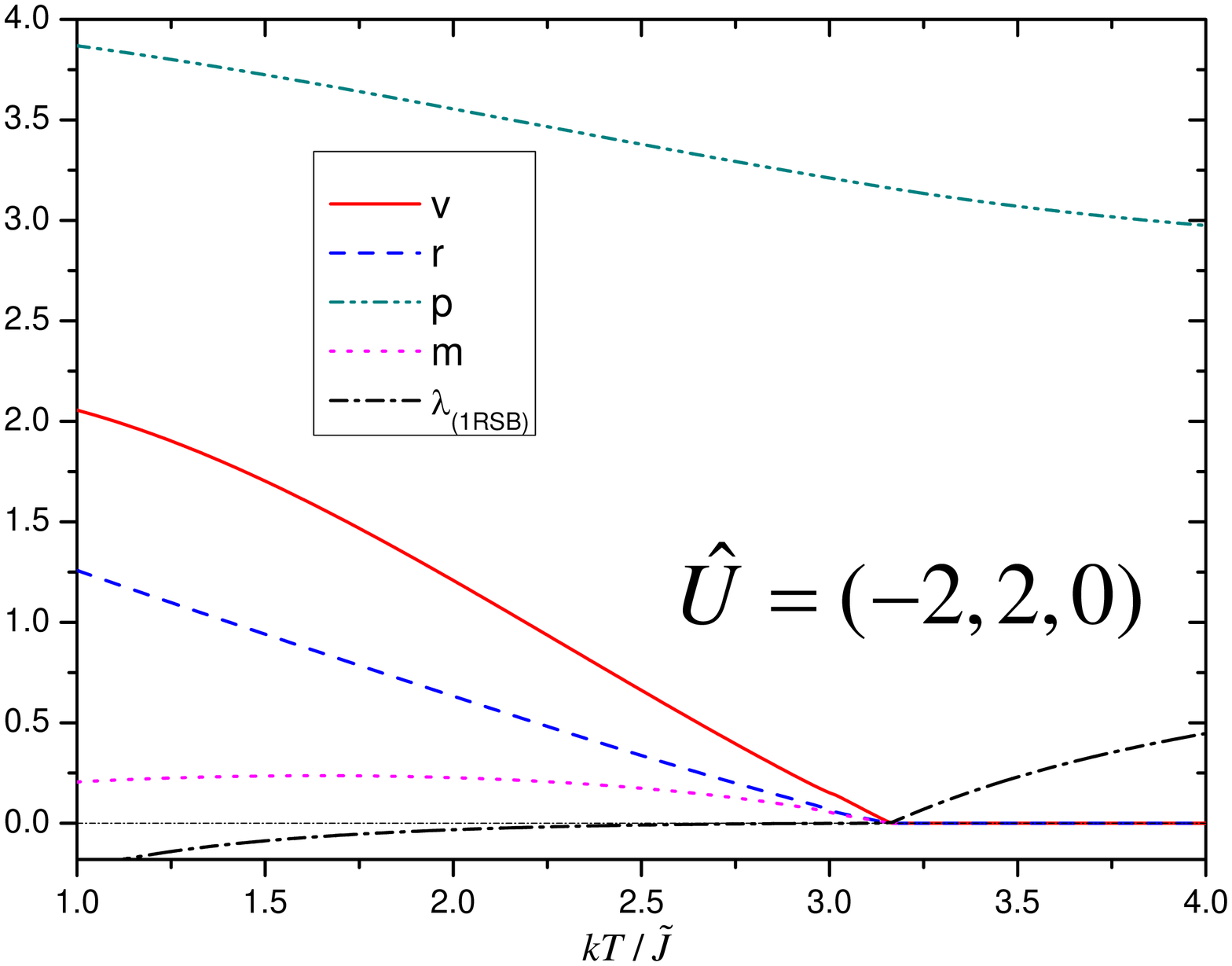}
\caption{1RSB order parameters and $\lambda_{\rm repl (1RSB)}$ for the case $\eta=1/\sqrt 3$
(analogue for spin-one glass model). Curves for $p_{\rm RS}$ and $p_{\rm 1RSB}$ coincide.}
\label{fig1}
\end{center}
\end{figure}
\begin{figure}[t]
\begin{center}
\includegraphics[height=100mm]{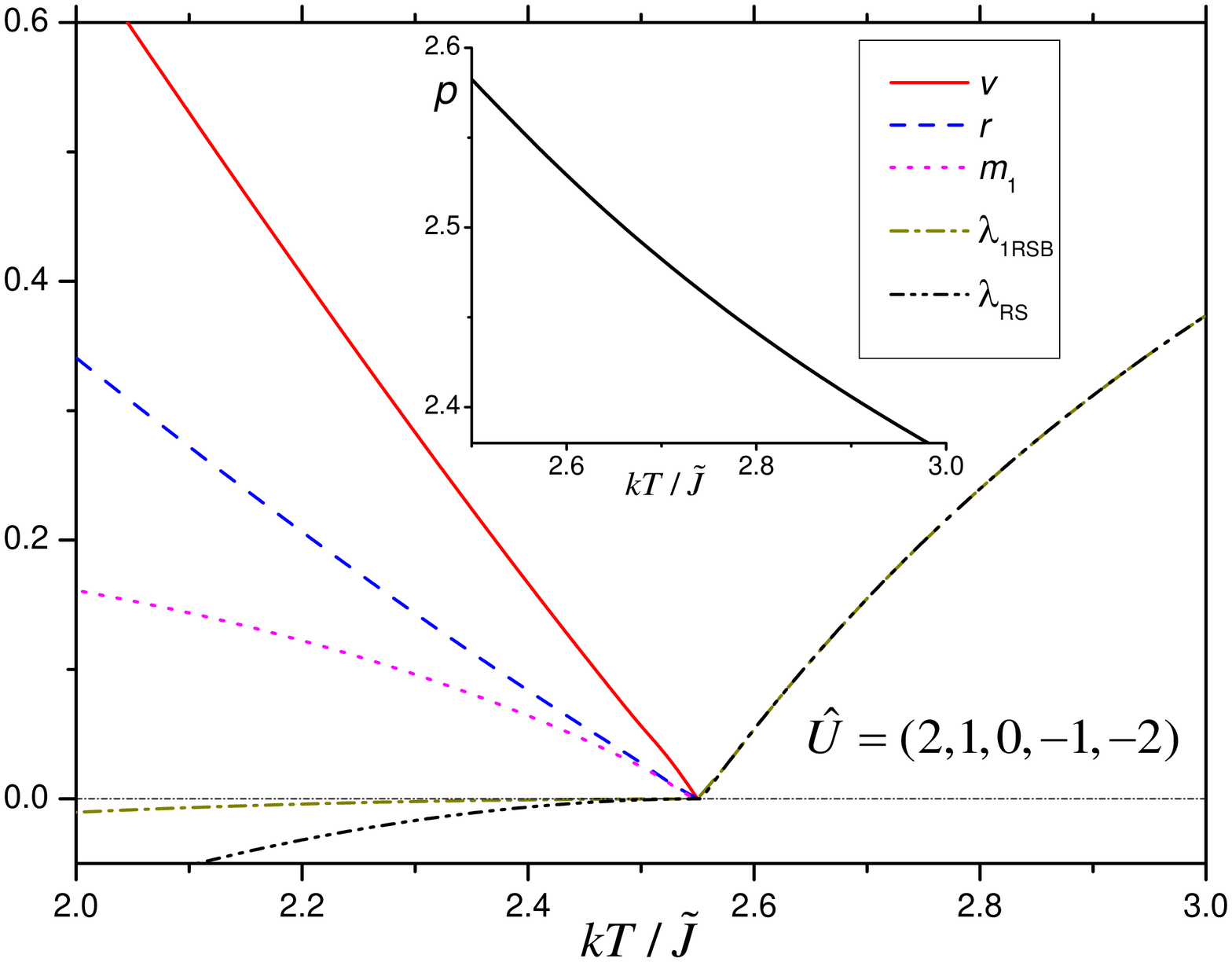}
\caption{1RSB order parameters, $\lambda_{\rm repl (RS)}$ and $\lambda_{\rm repl (1RSB)}$ for
spin-2 glass model. Curves for $p_{\rm RS}$ and $p_{\rm 1RSB}$ coincide.} \label{fig2}
\end{center}
\end{figure}

 In general case in FRSB scheme the series expansion in small $\delta
 q^{\alpha \beta }$ (up to fourth order) of the free energy ~(\ref{free})
 in the vicinity of $T_c$
 has the following form:
\begin{multline}
\frac {\Delta F}{NkT}=\lim_{n \rightarrow 0}\frac{1}{n} \left\{ \frac{t^2}{4}\sum (\delta
q^{\alpha \beta })^{2} - \frac{t^4}{4}\sum (\delta q^{\alpha\beta})^{2} \langle[U^2]\rangle^2-
\right.
\\
\frac{t^6}{6}\sum \delta q^{\alpha\beta} \delta q^{\beta\gamma } \delta q^{\gamma \alpha}
\langle[U^2]\rangle^3 - \\
\left.\frac{t^8}{8}\sum (\delta q^{\alpha\beta})^4 \left[\frac{3}{2} \langle[U^2]\rangle^4 -
\langle[U^4]\rangle\langle[U^2]\rangle^2+ \frac{1}{6} \langle[U^4]\rangle^2 \right]- \right.
\\
\frac{t^8}{8}\sum \delta q^{\alpha\beta} \delta q^{\beta\gamma } \delta q^{\gamma \delta }
\delta q^{\delta \alpha } \langle[U^2]\rangle^4+
\\
\left. \frac{t^8}{8}\sum (\delta q^{\alpha\beta})^2 (\delta q^{\alpha \gamma })^2 \left[3
\langle[U^2]\rangle^4 - \langle[U^4]\rangle\langle[U^2]\rangle^2 \right]\right\} \label{delta}
\end{multline}
Here
$$\hat{\theta}_0 = \frac{t^2}{2} p_c \sum (\hat {U}^{\alpha })^2$$
we use the notation
$$\left(\frac
{\Tr \hat {U}^l \exp \hat {\theta}_0}{\Tr \exp \hat {\theta}_0}\right)= \langle[U^l]\rangle$$
and
$$\hat{\theta}_0 = \frac{t^2}{2} p_c \sum (\hat {U}^{\alpha })^2.$$
The indices in ~(\ref{delta}) are different only if they are connected through $\delta
q^{\alpha \beta }$. We believe that the parameter $p$ ~(\ref{five}) changes slowly and is
equal to $p_c$ in the vicinity of $t_c$.

Now using Parisi rules
 ~\cite{par1,par2} we can write the free energy ~(\ref{delta}) in the
continuous case of FRSB in the form:
\begin{multline}
\frac{\Delta F}{NkT} = \frac{t^4}{4} \langle[U^2]\rangle^2 \langle q^2 \rangle(2\tau - \tau
^2)-
\\
\frac{t^6}{6} \langle[U^2]\rangle^3 \left[\int_0^1 dx x q^3 (x) + 3\int_0^1 dx q^2
(x)\int_x^1q(y) dy\right]+
\\
\frac{t^8}{48}\left[3 \langle[U^2]\rangle^2 - \langle[U^4]\rangle \right]^2 \langle
q^4\rangle-\frac{t^8}{8} \langle[U^2]\rangle^4 \left[\langle q^2\rangle^2 - 4\langle
q^2\rangle \langle q\rangle^2 - \right.
\\
\left. -4\langle q\rangle\int_0^1dx q(x) \int_0^x\left(q(y)-q(x)\right)^2 dy -\right.
\\
\left.\int_0^1dx \int_0^xdy \int_0^x dz \left(q(x)-q(y)\right)^2
\left(q(x)-q(z)\right)^2\right]-
\\
\frac{t^8}{8} \langle[U^2]\rangle^2 \left(3 \langle[U^2]\rangle^2 - \langle[U^4]\rangle
\right) \left[\langle q^4\rangle- \right.
\\
\left.2\langle q^2\rangle^2 - \int_0^1dx\int_0^xdy\left(q^2(x)-q^2(y)\right)^2\right] + ...
\label{contfree}
\end{multline}
where $$\langle q^l\rangle=\int_0^1dx q^l (x), ~~~~~ \tau =1-\frac{T}{T_c}, ~~~~~
\frac{kT_c}{\tilde J}=p(T_c).$$

The usual approach to find the extremum of the free energy with variation
with respect to $q(x)$ and further differentiations with respect to $x$
leads in the main approximation to
\begin{equation}
q(x) = \left\{%
\begin{array}{ll}
    cx, & \hbox{$0\leq x\leq\frac{\tau }{ct_c}$;} \\
    q(x)=\frac{\tau }{t_c}, & \hbox{$\frac{\tau }{c t_c}\leq x \leq 1$.} \\
\end{array}%
\right.
 \label{par}
\end{equation}
with
\begin{equation}
c=\frac{2\langle[U^2]\rangle^5}{\left[3\langle[U^2]\rangle^2 - \langle[U^4]\rangle\right]^2}
\label{par1}
\end{equation}
and $<q>=\tau /t_c$.

In close analogy to the procedure described in ~\cite{criri,crilepa} one can easily obtain the
expansion of the free energy and of  the order parameter $q(x)$ up to higher orders. One can
use the approach of Parisi equation~\cite{par1} for obtaining numerical solution. In our model
the free energy in an external field $h$ can be written in the form
\begin{equation}\frac{F(g,h)}{NkT}=\frac{t^2}{4}\left[p^2-\int_0^1q^2(x)dx
\right]-
\int_{-\infty}^{\infty}\frac{dy}{\sqrt
 {2 \pi q(0)}} \exp\left[-\frac{(y-h)^2}{2q(0)}\right]\phi (0,y),
\label{par2}
\end{equation}
where $\phi (0,y)$ is the solution at $x=0$ of the Parisi equation
\begin{equation}
{\dot \phi }(x,y)= - \frac {{\dot q}(x)}{2} \left[\phi '' (x,y)+x
(\phi '(x,y))^2\right]
\label{par3}
\end{equation}
with the boundary condition
\begin{equation}
\phi (1,y) = \ln \Tr \left\{\exp \left[ty {\hat U} + \frac {t^2}{2}
\left(p-q(1)\right) {\hat U}^2\right ] \right  \}.
\label{par4}
\end{equation}
The derivative with respect to $x$ are denoted by overdots and the derivatives with respect to
$y$ by primes. In obtaining ~(\ref{par3}) - ~(\ref{par4}) the fact that $\Tr \,\hat U^{(2k+1)}
=0$ was not used. However we assume that the FRSB solution exists and $q(0)=0$.

The results ~(\ref{contfree})-~(\ref{par4}) reduce to well known form
in the case of Sherrington--Kirkpatrick model \cite{sk} when $p=S^2=1$.
As another example the case of spin-one can be considered.
The replica symmetric solution for this model was given in
~\cite{spin1}. In this case $3p = 2+Q$ where $Q$ is the quadrupole moment.
The order--parameter function in FRSB scheme was obtained for spin--one
 model in ~\cite{costa} ($K=0$). Our results ~(\ref{par}) - ~(\ref{par1})
 can be reduced to those of ~\cite{costa}.

 Now we should like to emphasize the fact that if the Hamiltonian of a
 model is the sum of different Hamiltonians, each of them possessing
 reflection symmetry, the behavior of the system can be completely
 different relative to that discussed here in the case of nonzero
 mixed terms of third order. This can be demonstrated using the toy--model
 with the Hamiltonian
\begin{equation}
H=-\frac{1}{2}\sum_{i\neq j}J_{ij}\left[ \hat{U}^{(1)}_i\hat{U}^{(1)}_j +
\hat{U}^{(2)}_i\hat{U}^{(2)}_j + \hat{U}^{(3)}_i\hat{U}^{(3)}_j\right]
 \label{ham}
\end{equation}
where
 $J_{ij}$ are quenched Gaussian interactions
 with zero mean and
$$
\hat{U^{(1)}} = \left(
\begin{array}{ccc}
0& \phantom{x}0& \phantom{x}0\\
0& \phantom{x}1& \phantom{x}0\\
0& \phantom{x}0& -1\\
\end{array}
\right)
\hspace{7mm} \hat{U}^{(2)} = \left(
\begin{array}{cccc}
-1& \phantom{x}0& \phantom{x}0\\
\phantom{x}0& \phantom{x} 0& \phantom{x}0\\
\phantom{x}0& \phantom{x} 0& \phantom{x}1
 \end{array}
\right) \hspace{7mm} \hat{U}^{(3)} = \left(
\begin{array}{cccc}
1& \phantom{x} 0& \phantom{x}0\\
0&  -1& \phantom{x}0\\
0& \phantom{x} 0& \phantom{x}0
 \end{array}
\right)
$$

$\Tr(\hat {U}^{(l)})^3 =0$, but  $\Tr[(\hat {U}^{(l)})^2 \hat {U}^{(k)}]\neq 0$ if
 $l \neq k$ and this results in the appearance of the term
 $\sum (\delta q^{\alpha \beta })^{3}$ in the free
 energy expansion ~\cite{rsb}. The model described by the Hamiltonian
 ~(\ref {ham}) is isomorphic to three state Potts model and has an
 1RSB solution which is stable in a certain temperature  interval
 as was shown in ~\cite{grpr}.

To conclude, in the present paper we have shown the possibility of
 Parisi full replica symmetry breaking for rather general model with
 reflection symmetry. The model is a generalization of SK model. We obtain
 the main term of the expansion of the FRSB solution for the spin glass
 order parameter near $T_c$. The Parisi equation is also written in
 the general case.

Authors thank V.N.Ryzhov for
helpful discussions and valuable comments.

This work was supported in part by the Russian Foundation for
Basic Research (Grants No.05-02-17621 (EET) and No. 05-02-17280 (TIS
 and NMC) and by NWO-RFBR grant No. 04-01-89005 (047.016.001)(EET and NMC).
%
%






\end{document}